\documentclass[pra,twocolumn,superscriptaddress]{revtex4-1}
\usepackage{graphicx}
\usepackage{dcolumn}
\usepackage{bm}
\usepackage{color}
\usepackage{multirow}

\begin{document}


\title{Magnetisms of spinor alkali and alkaline-earth atoms in optical lattices}

\author{Hui Tan}
\affiliation{Department of Physics, National University of Defense Technology, Changsha 410073, P. R. China}
\author{Jinsen Han}
\affiliation{Department of Physics, National University of Defense Technology, Changsha 410073, P. R. China}
\author{Jianmin Yuan}
\affiliation{Department of Physics, National University of Defense Technology, Changsha 410073, P. R. China}
\affiliation{Department of Physics, Graduate School of China Academy of Engineering Physics, Beijing 100193, P. R. China}
\author{Yongqiang Li}
\email{li\_yq@nudt.edu.cn}
\affiliation{Department of Physics, National University of Defense Technology, Changsha 410073, P. R. China}

\date{\today}

\begin{abstract}
We theoretically investigate zero-temperature magnetic ordering of mixtures of spin-1 (alkali atoms) and spin-0 (alkaline-earth atoms) bosons in a three-dimensional optical lattice. With the single-mode approximation for the spin-1 bosons, we obtain an effective Bose-Hubbard model for describing the heteronuclear mixtures in optical lattices. By controlling the interspecies interactions between alkali and alkaline-earth atoms, we map out complete phase diagrams of the system with both positive and negative spin-dependent interactions for spin-1 atoms, based on bosonic dynamical mean-field theory. We find that the spin-1 components feature ferromagnetic and nematic insulating phases, in addition to the superfluid, depending on spin-dependent interactions. Take the spin-1 alkali bosons as spin $\uparrow$, and spin-0 alkaline-earth bosons as spin $\downarrow$, we observe that the system favors ferromagnetic insulator at filling $n=1$, and unorder insulator at $n=2$. Interestingly, we observe a two-step Mott-insulating-superfluid phase transition, as a result of mass imbalance between alkali and alkaline-earth atoms. 

\end{abstract}

\maketitle


\section{\label{sec:level1}introduction}
Quantum magnetism plays an important role in solid state system, and many theoretical and experimental efforts have been devoted to revealing the mechanisms behind magnetic ordering of quantum many-body systems~\cite{SachdevQuantum}. Most of studies in complex solid-state systems focus on magnetism of fermions~\cite{blundell2003magnetism}, since the basic element is electron. An exception in condensed matter physics is $^4$He, which is a bosonic system but its spin $S=0$. Magnetism of spinful bosonic systems is exclusive, even though bosonic magnetism can enrich our understanding of many-body
physics, especially about their quantum fluctuations. Therefore, it is desirable to study multi-species bosonic systems which are able to extend our understanding of quantum magnetism in many-body systems.

In the past decades, ultracold atoms trapped in periodic optical lattices have been utilized to study quantum many-body physics in a highly controllable manner~\cite{RevModPhys.80.885,RN29}, where multi-component ultracold gases composed of fermions or bosons have been achieved~\cite{jordens2008mott,schneider2008metallic}, opening new avenue for investigating quantum magnetism. Recently, an antiferromagnet has been realized in a repulsively interacting Fermi gas on a two-dimensional square lattice~\cite{MazurenkoA}. Two-component bosonic ultracold gases have also provided possibility for understanding quantum magnetism~\cite{PhysRevLett.106.195301,PhysRevLett.105.045303}, where a temperature in the order of picokelvin is achieved in a Bose-Bose mixture (pseudo-spin-1/2 Bose gases) in a three-dimensional optical lattice~\cite{PhysRevLett.106.195301}, even though the experimental temperature is still higher than the critical one of magnetic phase transition. For spin-1 bosons, the major experimental challenge is that the timescale of a spinor gas reaching ground states may exceed its lifetime, since the spin-dependent interaction (ferromagnetic or antiferromagnetic) is normally very small~\cite{RN36,RN38,RN39,RN40,RN41,RN42,RN43,PhysRevLett.87.010404,widera2006precision}.

Actually, the spinor bosonic gases lead to many interesting phenomena, including spin mixing~\cite{PhysRevA.60.1463}, spinwaves~\cite{PhysRevLett.89.090402}, spin dynamics~\cite{PhysRevLett.92.140403}, spin textures~\cite{PhysRevLett.100.170403}, and phase transitions~\cite{PhysRevA.86.061601,RN48,RN44,RN45,RN46,RN47}. Spinor bosonic gases in optical lattices provide possibilities of entering into strongly correlated regime, and normally demonstrate two quantum phases: Mott-insulating and superfluid phases with different types of long-range magnetic order, including nematic, cyclic, ferromagnetic and antiferromagnetic long-range order~\cite{RN26,RN28,RN25,RN27,RN29,RN26,RN30}. Recently, a phase transition from a longitudinal polar phase to a brokenaxisymmetry
phase in condensates of spin-1 ultracold sodium atoms is demonstrated  in a two-dimensional optical lattice~\cite{PhysRevLett.114.225302}, and thermal fluctuation induced stepwise Bose-Einstein condensation in a spinor gas is experimentally observed~\cite{PhysRevLett.119.050404}.

Moreover, heteronuclear mixtures of ultracold spinor bosons have been achieved experimentally, {\it i.e.} heteronuclear mixtures of spinor alkali atoms~\cite{PhysRevLett.114.255301}, or mixtures of spinor alkali and alkaline-earth atoms~\cite{barbe2018observation,RN51}, even though without a lattice. Different from identical spinor bosons with only even parity states being allowed for bosonic statistics, the heteronuclear mixtures of ultracold spin-1 alkali bosons in optical lattices demonstrate even richer competing many-body phases, which contains superfluid, spin-singlet, nematic, cyclic, charge-density-wave, and different types of ferromagnetic phases~\cite{luo2007bose,xu2009binary,LYQ2017}. However, magnetic phases of heteronuclear mixtures of spinor alkali and spin-0 alkaline-earth atoms in optical lattices are still unknown.

Motivated by the experimental work~\cite{barbe2018observation}, we investigate many-body ground states of heteronuclear mixtures of ultracold spin-1 ($^{87}$Rb or $^{23}$Na) and spin-0 alkaline-earth atoms ($^{84}$Sr) in a three-dimensional cubic optical lattice. To provide quantitative guidelines for upcoming experiments, we choose the experimental relevant parameters in our simulations, such as the ratio of hopping amplitudes between the two species, and of intra- and interspecies interactions. Here, we focus on quantum magnetism of the four-component system, and pay special attention to the influence of the spin-dependent interactions of spinor alkali atoms on quantum magnetism. To obtain the many-body ground states, we modify the recently developed four-component bosonic dynamical mean-field theory (BDMFT)~\cite{PhysRevLett.121.093401}, and extend the BDMFT equations for studying mixtures of spin-1 and spin-0 bosonic gases in optical lattices. Similar to fermionic cases~\cite{RN52,RN53,RN54}, we map the many-body lattice problem to a single-site problem which is coupled to two baths, $\it i.e.$ the condensed bath and the normal bath~\cite{RN54,RN55,RN56,RN57,RN58,RN59,RN60,RN61} to take quantum fluctuations into account.
In our work, complete phase diagrams of binary mixtures of ultracold spin-1 alkali and spin-0 alkaline-earth bosons are mapped out for different interspecies interactions. 

The paper is organized as follows: in section II we give a detailed derivation of the model and our approach. Section
III covers our results for heteronuclear mixtures of $^{87}$Rb and $^{84}$Sr, and of $^{23}$Na and $^{84}$Sr. We summarize with a discussion in Section V.

\section{Model and Method}
\subsection{Model}
We consider heteronuclear mixtures of spin-1 alkali atoms (such as $^{87}$Rb or $^{23}$Na with hyperfine spin $f=1$) and spin-0 alkaline-earth atoms ($^{84}$Sr with hyperfine spin $f=0$) in a three-dimensional (3D) cubic optical lattice. For a system of identical bosonic gases with hyperfine spin $f$, the general form of the interaction can be written in the second-quantized form~\cite{RN31,doi:10.1143/JPSJ.67.1822}:
\begin{equation}
\hat V(x_1-x_2)=\frac{4\pi\hbar^2}{m_a}\sum^{2f}_{F=0}a_F\hat P_F\delta(x_1-x_2),
\end{equation}
where $\hat P_F=\sum_{m_F}^{}\vert F,m_F\rangle\langle F,m_F\vert$ is the projection operator with $\vert F,m_F\rangle$ being the total hyperfine spin state formed by two atoms each with spin $f$, $m_a$ the atom mass, and $a_F$ the s-wave scattering length in the channel of total spin $F$.

The interaction for homonuclear mixtures of spin $f=1$ is given by:
\begin{eqnarray}
\hat V(x_1-x_2)&&=(g_0\hat P_0+g_2\hat P_2)\delta(x_1-x_2)\nonumber\\
& &=(c_0+c_2\bm{\hat S_1\cdot \hat S_2})\delta(x_1-x_2),
\end{eqnarray}
where $c_0=\frac{4\pi\hbar^2}{m_a}\frac{a_0+2a_2}{3}$, $c_2=\frac{4\pi\hbar^2}{m_a}\frac{a_2-a_0}{3}$,
and $\bm{\hat S_i}$ is the spin operator of the $i$th atom with spin-1. For heteronuclear mixtures of spin-0 and spin-1 atoms, the interaction between the two-species takes the form:
\begin{equation}
\hat V(x_1-x_2)=g_{12}\hat P_1\delta(x_1-x_2),
\end{equation}
where $g_{12}=\frac{2\pi\hbar^2a_{12}}{m_{12}}$ with the reduced mass $m_{12}=\frac{m_1m_2}{m_1+m_2}$. Here, $m_1$ and $m_2$ denote the atomic mass for species 1 and 2, respectively, and $a_{12}$ is the scattering length between spin-0 and spin-1 atoms. Note here that we assume the scattering lengths between spin-1 and spin-0 atoms are identical for the three-components of spin-1 atoms.

The many-body Hamiltonian for the system of heteronuclear mixtures of spin-0 and spin-1 bosonic gases takes the follow form:
\begin{widetext}
\begin{eqnarray}\label{Hamiltonian}
\hat{H}&=&\sum_{\nu,\sigma}\int dx\hat{\Phi}_{\nu_\sigma}^\dagger(x)(-\frac{\hbar^2\nabla^2}{2m_\nu}+V_{\rm lat})\hat{\Phi}_{\nu_\sigma}(x)
\nonumber\\
& &+\sum_{\sigma,\sigma'}\int dx\left[\frac{c_{0}}{2}\hat{\Phi}_{1_\sigma}^\dagger(x)\hat{\Phi}_{1_{\sigma'}}^\dagger(x)\hat{\Phi}_{1_{\sigma'}}(x)\hat{\Phi}_{1_\sigma}(x)+\frac{c_{2}}{2}\hat{\Phi}_{1_\sigma}^\dagger(x)\hat{\Phi}_{1_{\sigma'}}^\dagger(x)\hat{\bm{S}}_{\sigma\sigma'''}\cdot\hat{\bm{S}}_{\sigma'\sigma''}\hat{\Phi}_{1_{\sigma''}}(x)\hat{\Phi}_{1_{\sigma'''}}(x)\right]
\nonumber\\
& &+\int dx \frac {c^{(2)}}{2} \hat{\Phi}_{2_0}^\dagger(x)\hat{\Phi}_{2_0}^\dagger(x)\hat{\Phi}_{2_0}(x)\hat{\Phi}_{2_0}(x)+\sum_{\sigma}\int dxg_{12}\hat{\Phi}_{2_0}^\dagger(x)\hat{\Phi}^\dagger_{1_\sigma}(x)\hat{\Phi}_{1_\sigma}(x)\hat{\Phi}_{2_0}(x)
\end{eqnarray}
\end{widetext}
where $\hat{\Phi}_{\nu_\sigma}(x)$ is the field annihilation operator for the $\nu$th species ($\nu=1$ denoting spin-1 atoms, and $2$ spin-0 atoms) in the hyperfine state $\sigma=0,\,\pm1$ at point $x$, $V_{\rm lat}$ the optical lattice potential, and $c^{(2)}={4\pi\hbar^2a^{(2)}_0}/{m_2}$ with $a_0^{(2)}$ denoting the s-wave scattering length for species $\nu=2$ (only one component is considered for species 2, which is denoted as $\sigma\equiv0$ ) in the total spin $s=0$ channel.

By assuming a deep optical lattice potential and the single-mode approximation for spin-1 atoms~\cite{RN34}, we can expand the field operator by considering only the lowest energy band $\hat{\Phi}_{\nu_\sigma}(x)=\sum_{i}^{}\hat{b}_{i,\nu_\sigma}\omega_{\nu}(x-x_i)$, where the Wannier function of the lowest energy band $\omega_{\nu}(x-x_i)$ is well localized in the $i$th lattice site. Following the standard derivation for ultracold bosonic gases,  Eq.~(\ref{Hamiltonian}) reduces to an extended Bose-Hubbard model for heteronuclear mixtures of spin-0 and spin-1 bosons in an optical lattice, which can be written as:
\begin{widetext}
\begin{eqnarray}\label{Hubbard}
\hat{H}&=&-\sum_{\langle ij\rangle,\nu,\sigma}t_{\nu_\sigma}(\hat{b}_{i,\nu_\sigma}^\dagger \hat{b}_{j,\nu_\sigma}+{\rm H.c.})+\sum_{i,\sigma}U_{12} \hat{b}_{i,2_0}^\dagger \hat{b}_{i,1_\sigma}^\dagger \hat{b}_{i,1_\sigma}\hat{b}_{i,2_0}
\nonumber\\
& &+\sum_{i}
\left[
\frac{1}{2}U_1\hat{n}_{i,1}(\hat{n}_{i,1}-1)+\frac{1}{2}U'_1(\hat{\bm{S}}_{i,1}^2-2\hat{n}_{i,1})+\frac{1}{2}U_2\hat{n}_{i,2}(\hat{n}_{i,2}-1)-\mu_{1}\hat{n}_{i,1}-\mu_{2}\hat{n}_{i,2}
\right],
\end{eqnarray}
\end{widetext}
where $\hat{b}_{i,\nu_\sigma}^\dagger(\hat{b}_{i,\nu_\sigma})$ is the bosonic creation (annihilation) operator for species $\nu$ and hyperfine state $\sigma$ at site $i$, $\hat{n}_{i,\nu}=\sum_{\sigma}\hat{n}_{i,\nu_\sigma}$ with $\hat{n}_{i,\nu_\sigma}=\hat{b}_{i,\nu_\sigma}^\dagger \hat{b}_{i,\nu_\sigma}$ being the number of particles, $\hat{\bm{S}}_{i,1}=\sum_{\sigma\sigma'}\hat{b}_{i,\nu_\sigma}^\dagger
{\bf\Gamma}_{\sigma\sigma'} \hat{b}_{i,\nu_{\sigma'}}$ is the local spin operator with ${\bf\Gamma}_{\sigma\sigma'}$ being the usual spin matrices for a spin-1 particle, $\mu_\nu$ denotes the chemical potential for species $\nu$, and $t_{\nu_\sigma}$ the hopping amplitude on the lattice where only hopping between nearest neighbors $\langle ij \rangle$ is considered. The Hubbard repulsion $U_1=c_0\int d\bm r\vert\omega_1(\bm r-\bm r_i)\vert^4$ and spin-dependent interaction $U_1'=c_2 \int d\bm r\vert\omega_{1}(\bm r-\bm r_i)\vert^4$ for spin-1 bosons, $U_2=c^{(2)}\int d\bm r\vert\omega_2(\bm r-\bm r_i)\vert^4$ for spin-0 bosons, and interspecies interaction $U_{12} = g_{12} \int d\bm r\vert\omega_1(\bm r-\bm r_i)\omega_2(\bm r-\bm r_i)\vert^2$ between spin-1 and spin-0 bosons. Note here that only one component is considered for species 2 with $\sigma=0$. 

\subsection{Theoretical method}
To investigate quantum phenomena and obtain many-body ground states of binary mixtures of spinor alkali and alkaline-earth Boses loaded into a three-dimensional cubic optical lattice, we utilize an extended bosonic version of dynamical mean-field theory (BDMFT) to solve the problem  described by Eq.~(\ref{Hubbard}). The main idea of the BDMFT approach is to map the quantum lattice problem with many degrees of freedom onto a single-site problem, which is coupled self-consistently to two noninteracting baths~\cite{PhysRevLett.121.093401,RN54}. Here, one bath is a condensed one, and another a normal one. BDMFT treats the condensed and normal bosons on equal footing, and expands the equations up to second order as a function of $1/z$, where BDMFT takes the lattice coordination number $z$ as the control parameter ($z=6$ for three-dimensional cubic lattice). 

Even though the general processes of the BDMFT derivations here are similar to spin-1/2 bosons in $p$-band optical lattices~\cite{PhysRevLett.121.093401}, the hopping amplitudes and interactions for mixtures of spin-1 and spin-0 bosons are different, leading to totally different Green's functions and self energies. Following the standard derivation for BDMFT~\cite{RN54,RN55}, we can write down the BDMFT equations in an explicit form, where the corresponding effective action of the impurity site is described by:
\begin{widetext}
\begin{eqnarray}\label{effictive_interaction}
S_{\rm imp}^{(0)}&=&-\int_{0}^{\beta}d\tau d\tau'\sum_{\nu_\sigma, \nu_{\sigma'}}
\left(
\begin{array}{cc}
b_{\nu_\sigma}^{(0)*}(\tau)&b_{\nu_\sigma}^{(0)}(\tau)
\end{array}
\right)
{\mathcal{G}_{\sigma\sigma'}^{(0)}}^{-1}(\tau-\tau')
\left(
\begin{array}{c}
b_{\nu_{\sigma'}}^{(0)}(\tau')\\
b_{\nu_{\sigma'}}^{(0)*}(\tau')
\end{array}
\right)+\int_{0}^{\beta}d\tau U_{12}n_1^{(0)}(\tau)n_2^{(0)}(\tau)\\
& &+\int_{0}^{\beta}d\tau
\left\{
\frac{1}{2}\sum_\nu U_\nu n_\nu^{(0)}(\tau)(n_\nu^{(0)}(\tau)-1)+\frac{1}{2}U'_1\left(\bm S_1^{(0)}(\tau)^2-2n_1^{(0)}(\tau)\right)-\sum_{\langle 0i\rangle,\nu_\sigma}t_{\nu_\sigma}\left(b_{\nu_\sigma}^{(0)*}(\tau)\phi_{i,\nu_\sigma }^{(0)}(\tau)+{\rm H.c.}\right)\nonumber
\right\}
\end{eqnarray}
\end{widetext}

Here, ${\mathcal{G}_{\sigma\sigma'}^{(0)}}(\tau-\tau')$ is the non-interacting Weiss Green's function:
\begin{widetext}
	\begin{eqnarray}
	{\mathcal{G}_{\sigma\sigma'}^{(0)}}^{-1}(\tau-\tau')=-
	\left(
	\begin{array}{cc}
	(\partial_{\tau'}-\mu_{\sigma})\delta_{\sigma\sigma'}+t^2\sum_{\langle 0i\rangle,\langle 0j\rangle}G_{\sigma\sigma',ij}^1(\tau,\tau')&t^2\sum_{\langle 0i\rangle,\langle 0j\rangle}G_{\sigma\sigma',ij}^2(\tau,\tau')\\
	t^2\sum_{\langle 0i\rangle,\langle 0j\rangle}{G_{\sigma\sigma',ij}^2}^*(\tau',\tau)&(-\partial_{\tau'}-\mu_{\sigma})\delta_{\sigma\sigma'}+t^2\sum_{\langle 0i\rangle,\langle 0j\rangle}G_{\sigma\sigma',ij}^1(\tau',\tau)
	\end{array}
	\right),	
	\end{eqnarray}
\end{widetext}
which is a $8\times 8$ matrix with $\sigma$ running over all the possible values $\sigma=0$, $\pm 1$ for species 1, and $\sigma=0$ for species $2$, to shorten the notation of Green's functions. Explicitly, the diagonal- and off-diagonal parts of Green's functions read:
\begin{eqnarray}
&&G_{\sigma\sigma',ij}^1(\tau,\tau') \equiv-\langle b_{i,\sigma}(\tau)b_{j,\sigma'}^*(\tau')\rangle_0+\phi_{i,\sigma'}(\tau)\phi_{j,\sigma}^*(\tau'),\nonumber\\
&&G_{\sigma\sigma',ij}^2(\tau,\tau') \equiv-\langle b_{i,\sigma}(\tau)b_{j,\sigma'}(\tau')\rangle_0+\phi_{i,\sigma'}(\tau)\phi_{j,\sigma}(\tau'),\nonumber
\end{eqnarray}
where $\phi_{i,\sigma}(\tau)\equiv\langle b_{i,\sigma}(\tau)\rangle_0$ is introduced as the superfluid order parameters ($\sigma=0,\,\pm 1$ for species 1, and $\sigma=0$ for species 2), and in the cavity system $\langle \dots\rangle_0$ denotes the expected value without the impurity site.

Normally, it is difficult to find an analytical solver for the effective action. To obtain many-body ground states, we turn back to the Hamiltonian representation and represent the effective action described in Eq.~(\ref{effictive_interaction}) by the Anderson impurity Hamiltonian:
\begin{widetext}
\begin{eqnarray}
\hat{H}_A^{(0)}&=&\sum_{\nu,\sigma}
\left(
-t_{\nu_\sigma}\left(\phi^{*(0)}_{\nu_\sigma}\hat{b}_{\nu_\sigma}^{(0)}+{\rm H.c.}\right)+\frac{1}{2}U_{1}\hat{n}_{1}^{(0)}\left(\hat{n}_{1}^{(0)}-1\right)+\frac{1}{2}U'_{1}\left(\hat{\bm S}_{1}^{(0)2}-2\hat{n}_{1}^{(0)}\right) + \frac{1}{2}U_{2}\hat{n}_{2}^{(0)}\left(\hat{n}_{2}^{(0)}-1\right) -\mu_{\nu}\hat{n}_{\nu_\sigma}^{(0)}
\right)\nonumber\\
&&+U_{12}\hat{n}_1^{(0)}\hat{n}_2^{(0)}+\sum_{l}\epsilon_l\hat{a}_l^{\dagger}\hat{a}_l+\sum_{l,\nu,\sigma}\left(V_{\nu_\sigma,l}\hat{a}_l^{\dagger}\hat{b}_{\nu_\sigma}^{(0)}+W_{\nu_\sigma,l}\hat{a}_l\hat{b}_{\nu_\sigma}^{(0)}+ {\rm H.c.}\right),
\end{eqnarray}
\end{widetext}
where only one component is considered for species $\nu=2$, {\it i.e.} $\sigma\equiv0$ for $\nu=2$.
Obviously, the interaction terms are identical with that in the Hubbard Hamiltonian, as in Eq.~(\ref{Hubbard}), since all the interactions considered here are local ones. As we mentioned above, BDMFT has reduced the many-body lattice problem to a single-site problem coupled to the condensed and normal baths, and presented equations up to subleading order. Here, the leading term is the Gutzwiler term with order parameters $\phi_{\nu_\sigma}$ standing for the condensed bath, and the subleading term is the normal bath described by operators $\hat{a}_l^{\dagger}$ with energies $\epsilon_l$, where the coupling between the normal bath and impurity site is realized by $V_{\nu_\sigma,l}$ (normal-hopping amplitudes) and $W_{\nu_\sigma,l}$ (anomalous-hopping amplitudes).
And then, the Anderson impurity model can be solved through numerical methods, and detailed steps have been introduced in Ref.~\cite{RN48,RN49}. Here, we use exact diagonalization as the solver to obtain the many-body ground states. We remark here that BDMFT is nonperturbative, and hence can be applied within the full range from small to large couplings.

\section{results}
\begin{figure}
\includegraphics[width=1\linewidth]{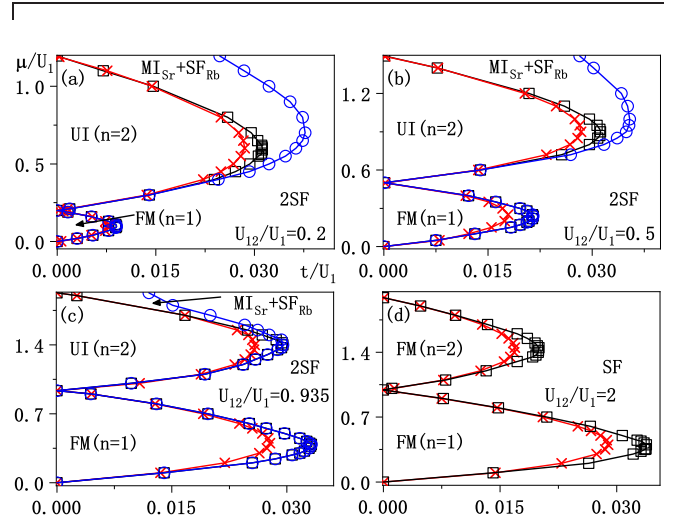}
\caption{\label{phase_diagram} Phase diagrams of heteronuclear mixtures of ultracold spin-1 $^{87}$Rb (spin-dependent interaction $U'_1/U_1=-0.0046$) and spin-0 $^{84}$Sr bosons in a 3D cubic lattice for different interspecies interactions $U_{12}/U_1=0.2,0.5,0.935$ (scattering length from Ref.~\cite{barbe2018observation}) and $2$, obtained by BDMFT. The system favors ferromagnetic insulating phase (FM) at filling $n=1$, unorder insulating phase (UI) at $n=2$, and two types of superfluid (MI$_{\rm Sr}$ + SF$_{\rm Rb}$, and 2SF), where the three-components of spin-1 $^{87}$Rb demonstrate ferromagnetic order as a result of ferromagnetic interactions. Note here that the system favors phase separation for $U_{12}/U_1=2$, and here we only show the phase diagram of spin-1 bosons. For comparisons, the red cross is obtained by Gutzwiller mean-field theory. The other parameters $t \equiv t_{1_\sigma}\approx 0.97t_{2_0}$, and $U_2/U_1=1.26$.}
\end{figure}

\begin{figure}
	\centering
	\includegraphics[width=1\linewidth]{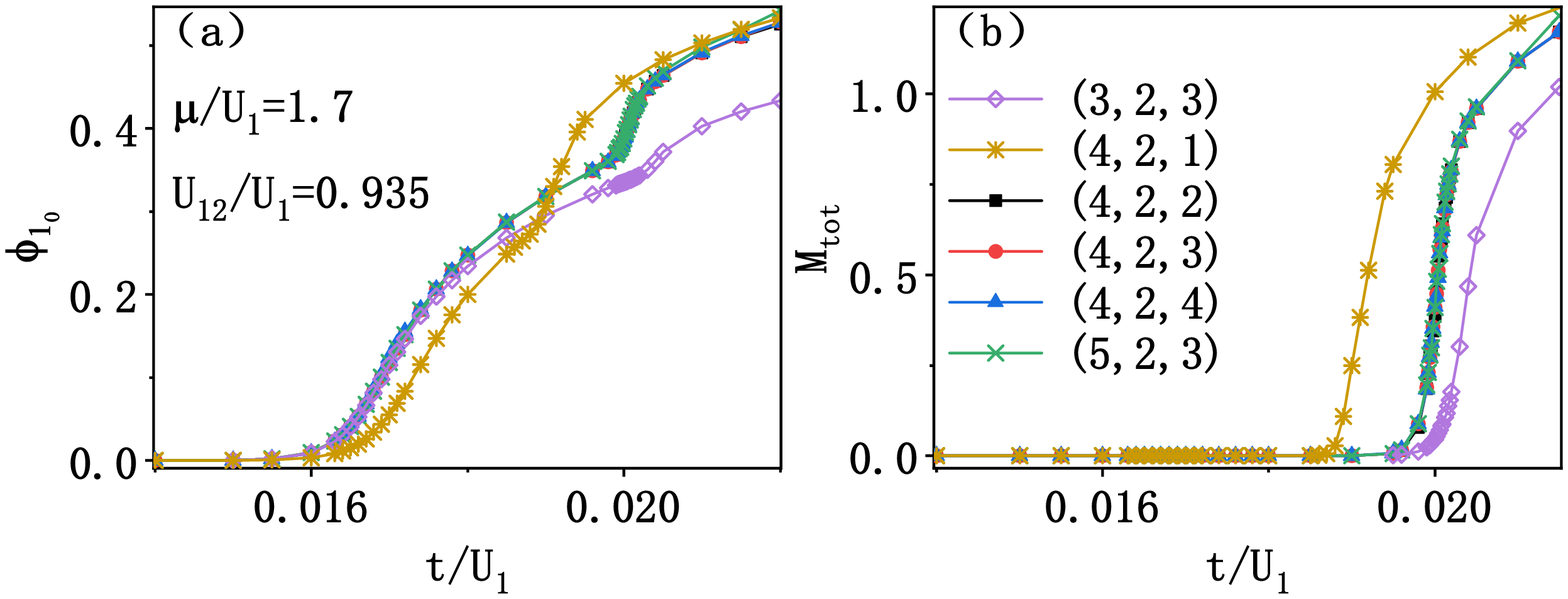}
	\caption{\label{bath} Order parameters and magnetism as a function of hopping amplitudes under different parameter $(N,N_{orb},L)$, where the results are converged for higher truncation values. Here, $N$ is the maximum occupation number truncated for the Fock space, $N_{orb}$ the maximum occupation number of the orbital for each normal bath, and $L$ the number of normal bath orbitals. The chemical potential $\mu/U_1=1.7$, interactions $U_{12}/U_1=0.935$, $U_{2}/U_1=1.26$, $U^\prime_1/U_1=-0.0046$, and hopping amplitudes $t\equiv t_{1_\sigma}\approx 0.97t_{2_0}$.}
\end{figure}

In this paper, we investigate phase diagrams of heteronuclear mixtures of spin-1 and spin-0 atoms in a 3D optical lattice, which is characterized by the order parameter $\phi_{\nu_\sigma}=\langle \hat{b}_{\nu_\sigma} \rangle$, magnetism ${ \hat{\bf S}}_{1}=\hat{b}_{1_\sigma}^\dagger{\bf\Gamma}_{\sigma\sigma'} \hat{b}_{1_{\sigma'}}$ for spin-1 bosons, {\it i.e.} $ \hat{S}_{1_x}=1/\sqrt{2}(\hat{b}^\dagger_{1_{+1}}\hat{b}_{1_0}+\hat{b}^\dagger_{1_0}\hat{b}_{1_{+1}}+\hat{b}^\dagger_{1_0}\hat{b}_{1_{-1}}+\hat{b}^\dagger_{1_{-1}}\hat{b}_{1_0})$, $ \hat{S}_{1_y}=i/\sqrt{2}(-\hat{b}^\dagger_{1_{+1}}\hat{b}_{1_0}+\hat{b}^\dagger_{1_0}\hat{b}_{1_{+1}}-\hat{b}^\dagger_{1_0}\hat{b}_{1_{-1}}+\hat{b}^\dagger_{1_{-1}}\hat{b}_{1_0})$, and $\hat{S}_{1_z}=(\hat{b}^\dagger_{1_{+1}}\hat{b}_{1_{+1}}-\hat{b}^\dagger_{1_{-1}}\hat{b}_{1_{-1}})$, and local total magnetism of spin-1 and spin-0 atoms ${\hat{\bf S}} =\sum_{\sigma{\sigma'}} \hat{b}^\dagger_{\nu_{\sigma}} {\bf F}_{\nu_\sigma\nu'_{\sigma'}} \hat{b}_{\nu'_{\sigma'}} $,
with ${\bf F}_{\nu_\sigma\nu^\prime_{\sigma'}}$ denoting the spin matrix for a spin-1/2 particle formed by $\sigma$-component of species 1 and $0$-component of species 2, {\it i.e.} $\hat{S}_x= 1/2  \sum_\sigma({\hat b}^\dagger_{1_\sigma} {\hat b}_{2_0} + {\hat b}^\dagger_{2_0} {\hat b}_{1_\sigma}) $, $\hat{S}_y=i/2 \sum_\sigma (-{\hat b}^\dagger_{1_\sigma} {\hat b}_{2_0} + {\hat b}^\dagger_{2_0} {\hat b}_{1_\sigma}) $, and $\hat{S}_z= 1/2 \sum_\sigma ({\hat b}^\dagger_{1_\sigma} {\hat b}_{1_\sigma} - {\hat b}^\dagger_{2_0}  {\hat b}_{2_0} )$. Here, we treat the spin-1 alkali atoms as spin $\uparrow$, and spin-0 alkaline-earth atoms as spin $\downarrow$, and essentially achieve a pseudo-spin-1/2 bosonic model composed of spin-0 atoms and each component of spin-1 atoms in optical lattice, since only density-density interactions appear between spin-0 and each component of spin-1 atoms. Therefore, we focus on correlations between Sr and each component of spin-1 Rb or Na. In our calculations, we consider the spin-miscible regime with inter-species interactions $U_{12}< \sqrt{U_1 U_2}$, and pay special attention to the lower filling cases. We focus on both the ferromagnetic $U'_1/U_1<0$ ($^{87} {\rm Rb}$) and antiferromagnetic interactions $U'_1/U_1>0$ ($^{23} {\rm Na}$). In all our simulations, we set $U_1\equiv 1$ as the unit of energy.

\begin{figure*}
	\centering
	\includegraphics[width=1\linewidth]{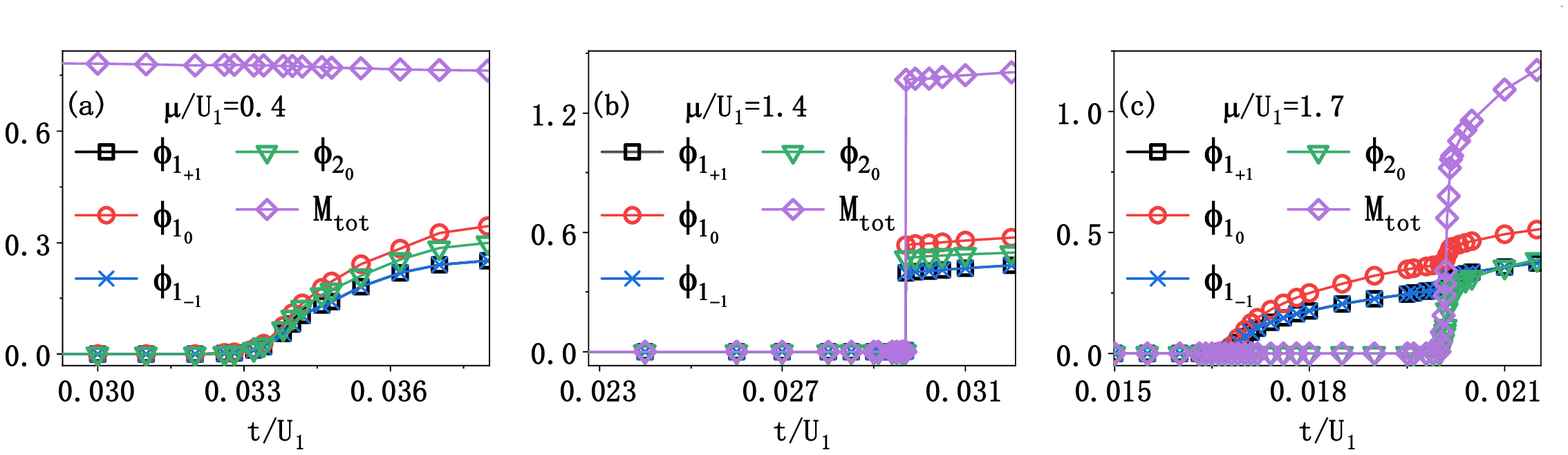}
	\caption{\label{phi_n} Zero-temperature phase transitions for mixtures of ultracold spin-1 $^{87}$Rb and spin-0 $^{84}$Sr bosons in a 3D cubic lattice, where order parameters $\phi_{\nu_\sigma}$ and local magnetism $M_{\rm tot}$ are shown. We observe a second-order phase transition for chemical potential $\mu /U_1=0.4$ ($n=1$ at Mott phase), a first-order phase transition at $\mu=1.4$ ($n=2$ at Mott regime), and a two-step phase transition at $\mu/U_1=1.7$ ($n=2$ at Mott regime). Here, we choose the experimental relevant parameters with interactions $U_{12}/U_1=0.935$, $U_{2}/U_1=1.26$, and hopping amplitudes $t\equiv t_{1_\sigma}\approx 0.97t_{2_0}$.}
\end{figure*}

\subsection{Mixtures of spinor $^{87} {\rm Rb}$ and $^{84} {\rm Sr}$}
We first study mixtures of spin-1 $^{87} {\rm Rb}$ and spin-0 $^{84} {\rm Sr}$ atoms in a three-dimensional optical lattice, where the ${\rm Rb}$ atoms possess a ferromagnetic interaction $U'_1/U_1=-0.0046$~\cite{RN62}, interaction between the ${\rm Sr}$ atoms $U_2/U_1=1.26$~\cite{RN63}, and interspecies interaction between the ${\rm Rb}$ and ${\rm Sr}$ atoms $U_{12}/U_1\approx0.935$, as achieved in experiments~\cite{barbe2018observation,RN51}.
In out simulations, we choose the hopping amplitudes $t \equiv t_{1_\sigma} \approx 0.97 t_{2_0}$, due to the mass relation $t_{1_\sigma}/t_{2_0}=m_2/m_1$, and the chemical potential $\mu \equiv \mu_{1}=\mu_2$.

Our results are summarized in the Fig.~\ref{phase_diagram}, where we map out the low-filling lobes under different interspecies interactions $U_{12}/U_1=$ 0.2, 0.5, 0.935, and 2, obtained by bosonic dynamical mean-field theory. We observe there are four phases in the low-filling regime, including ferromagnetic phase (FM) characterized by $M_{\rm tot}\equiv\vert \langle { \hat{\bf S}}\rangle\vert \ne 0$ and $\phi_{1_\sigma}=\phi_{2_0} = 0 $, unorder insulator (UI) characterized by $M_{\rm tot}= 0$ and $\phi_{1_\sigma}=\phi_{2_0}=0 $, two types of superfluid phases (SF). In the lower hopping regime $t\ll U_1$, the system demonstrates Mott-insulating phases with the spin-1 Rb atoms favoring ferromagnetic spin order, which is characterized by $\phi_{1_\sigma}=0$ and $M_1=|\langle { \hat{\bf S}}_1\rangle| \ne 0$, and the physical reason is that, as expected, the ferromagnetic interaction $U^\prime_1<0$ supports ferromagnetic order to lower the energy of the spin-1 atoms~\cite{RN44,RN45}. The results also indicate that interspecies density-density interactions between Rb and Sr do not influence magnetic order of spin-1 bosons. We remark here that the local spin $M_1$ of the $^{87}{\rm Rb}$ always meet the relationship $ M_1/n_{Rb} = 1$, with $n_{Rb}$ being the local total filling of $^{87}{\rm Rb}$. To characterize the properties of the whole system mixed by Rb and Sr, we define the local total magnetism by taking Rb as spin $\uparrow$ and Sr as spin $\downarrow$. We find that the system possesses nonzero magnetism $M_{\rm tot} \ne 0$ for filling $n\equiv n_{Rb}+n_{Sr}=1$ ($n_{Rb}=n_{Sr}=0.5$) with $n_{Sr}$ ($n_{Rb}$) denoting the local filling of $^{84}{\rm Sr}$ ($^{87}{\rm Rb}$), and zero magnetism $M_{\rm tot}=0$ for filling $n=2$, which indicates our system favor ferromagnetic insulating phase (FM) at filling $n=1$, and unorder insulating phase (UI) at $n=2$. To understand the long-range spin order defined here, we can utilize the effective spin model in the strongly interacting regime for Bose-Bose mixtures in optical lattices~\cite{altman2003phase}, and, here, the mixtures between each component of the spin-1 atoms and spin-0 atoms essentially yield a spin-1/2 model for each component of spin-1 atoms. The underlying physics is that, for filling $n=1$, ferromagnetic spin coupling dominates for the parameters with almost identical hopping amplitudes $t_{1_\sigma}\approx t_2$ and ferromagnetic long-range order develops, whereas, for filling $n=2$, spin fluctuations are suppressed for $U_{12}\ll U_{1,2}$ and an unorder insulating phase appears.
With the increase of the hopping amplitudes, density fluctuations dominate and superfluid phases appear with $\phi_{\nu_\sigma}\neq0$. Due to the mass imbalance, the Rb atoms with larger mass delocalize first (${\rm MI}_{\rm Sr}$ + ${\rm SF}_{\rm Rb}$), and then both species delocalize (2SF) with increasing hopping amplitudes.

As we have mentioned above, BDMFT reduce the many-body lattice problem to a single-site problem coupled to the condensed and normal baths. Thus, the convergence of the BDMFT method under different number of baths should be verified. As shown in Fig.~\ref{bath}, we plot the order parameter $\phi_{1_0}$ and local magnetism $M_{\rm tot}$ as a function of hopping amplitudes under different parameters $(N,N_{orb},L)$. Here, $N$ denotes the maximum occupation number which is used to truncate for the Fock space, $N_{orb}$ the maximum occupation number of the orbital for each normal bath, and $L$ the number of normal bath orbitals. We clearly observe that our results are converged for higher truncation values, and typically choose $N=4$, $N_{orb}=2$, $L=3$ in the our simulations in the low filling regime.

Next, we study the influence of interspecies interactions $U_{12}$ on the phase diagrams. For smaller interspecies interactions with $U_{12}<\sqrt{U_1U_2}$, the two species are miscible, and the system favors phase separation for $U_{12}>\sqrt{U_1U_2}$. As shown in Fig.~\ref{phase_diagram}(a-d), we observe that the first lobe with filling $n=1$ shrinks, and the second lobe with $n=2$ expands with decreasing $U_{12}$. The physical reason is that the spins $\uparrow$ and $\downarrow$ compose a spin singlet with $M_{\rm tot}=0$ for $n=2$ (as shown in Fig.~\ref{phi_n}), which is more favorable for smaller $U_{12}$. In the limit of $U_{12}\ll U_{1,2}$, we remark here that the phase diagrams of Rb and Sr almost coincide, if we scale the hopping amplitudes by their own onsite interactions, {\it i.e.} $t_{1_\sigma}/U_1$ and $t_{2_0}/U_2$, since the spin-dependent interaction of spin-1 Rb atoms is tiny. For a larger interspecies interaction $U_{12}/U_1=2$, we observe a phase separation in the system. Here, we only plot the phase diagram of spin-1 $^{87}{\rm Rb}$ atoms, and recover the phase diagram of spin-1 bosonic systems in an optical lattice, as shown in Fig.~\ref{phase_diagram}(d).

For comparisons, we also investigate the mixtures of spin-1 and spin-0 bosons in optical lattices, based on Gutzwiller mean-field theory, as shown by the red lines in Fig.~\ref{phase_diagram} (only the first Mott-insulating-superfluid transition is shown). As stated in the method part, Gutzwiller mean-field method is actually first-order approximation of BDMFT, and, by expanding to second order as a function of coordination number $z$, quantum fluctuations are taken into account and BDMFT is achieved. The phase diagrams clearly show that the tips of Mott lobe from Gutzwiller method are smaller, compared to the black lines obtained by BDMFT, as a result of quantum fluctuations which are included in BDMFT. We remark here that Gutzwiller method cannot resolve magnetic long-rang order of Mott phases of the bosonic systems in optical lattices, since quantum magnetism is attributed to second-order tunneling processes, which are neglected in Gutzwiller static mean-field theory.

\begin{figure}
	\centering
	\includegraphics[width=0.9\linewidth]{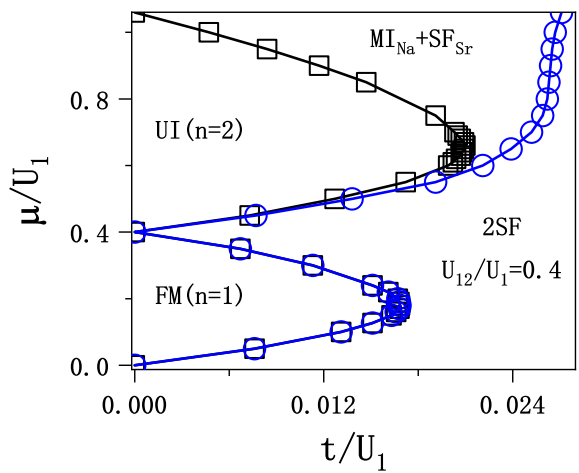}
	\caption{\label{Na_phase} Phase diagram of heteronuclear mixtures of ultracold spin-1 $^{23}$Na (antiferromagnitic interaction $U'_1/U_1=0.037$) and spin-0 $^{84}$Sr bosons in a 3D cubic lattice. The system supports four phases in the low-filling regime, including ferromagnetic phase (FM), unorder insulator (UI), two kinds of superfluid phases (MI$_{\rm Na}$ + SF$_{\rm Sr}$, and 2SF). The other parameter are $t \equiv t_{1_\sigma}= t_{2_0}$, $U_{2}/U_1=0.66$, and $U_{12}/U_1=0.4$.}
\end{figure}

To obtain phase boundaries of the heteronuclear mixtures of spin-1 and spin-0 bosons in Fig.~\ref{phase_diagram}, we plot the order parameters $\phi_{\nu_\sigma}$ and magnetism $M_{\rm tot}$ as a function of hopping amplitudes for different chemical potentials, as shown in Fig.~\ref{phi_n}. As well known, the phase transition is filling dependent for multicomponent bosonic mixtures in optical lattices. For example, the phase transition is second order for filling $n=1$, and can be first order for filling $n=2$ for Bose-Bose mixtures~\cite{RN58,YamamotoFirst,condmat5010002,PhysRevA.91.043615} and spin-1 bosons~\cite{RN47}. The physical reason for the first-order phase transition is that the system favors a spin-singlet Mott-insulating state with even filing, which is discontinuous with the superfluid phase with nonzero magnetism, indicating a jump of spin long-range order.
As shown in Fig.~\ref{phi_n}, we clearly observe a second order phase transition from Mott insulator to superfluid for filling $n=1$  [$\mu/U_1=1.4$, Fig.~\ref{phi_n}(a)], and a first-order phase transition for filling $n=2$ [$\mu/U_1=1.4$, Fig.~\ref{phi_n}(b)]. More interestingly, we also observe a two-step phase transition at larger chemical potential [$\mu/U_1=1.7$, Fig.~\ref{phi_n}(c)], $\it i.e.$ the Rb atoms delocalize first and then the Sr atoms, since the heavier Rb atoms induce $U_1<U_2$.

\subsection{Mixtures of spinor $^{23} {\rm Na} $ and $^{84} {\rm Sr}$}
In contrast to the results that we mentioned above, in this section we focus on the mixtures of spin-1 $^{23} {\rm Na}$ and spin-0 $^{84} {\rm Sr}$ in a 3D optical lattice. Different from spin-1 $^{87} {\rm Rb}$, the spin-dependent interaction here is an antiferromagnetic one for $^{23} {\rm Na}$ with $U'_1/U_1=0.037$~\cite{RN39}, and $U_{2}/U_1=0.66$~\cite{RN63}. As far as we know, there are no experimental data for heteronuclear mixtures of spinor $^{23} {\rm Na} $ and spin-0 $^{84} {\rm Sr}$, and, without loss of generality, we choose $t\equiv t_{1_\sigma}=t_{2_0}$, $\mu\equiv\mu_1=\mu_2$, and $U_{12}/U_{1}=0.4$ for a spin-miscible case.

For the spin-1 $^{23} {\rm Na}$ bosons with antiferromagnetic interactions, the spin-1 bosons favor different types of spin long-range order in the Mott-insulating region depending on the local filling. For example, the spin-1 bosons in optical lattices demonstrate spin nematic order for odd filling, characterized by $\phi_{1_\sigma}=0$, $\varphi_{1_{\alpha\beta}}=\langle \hat{S}_{1_\alpha}\hat{S}_{1_\beta} \rangle-\delta_{\alpha\beta}\cdot \langle { \hat{\bf S}}^2_1 \rangle /3>0$ and $M_1=0$, and spin-singlet order for even filling, characterized by $\phi_{1_\sigma}=0$,$\varphi_{1_{\alpha\beta}}=0$ and $M_1=0$, with $\alpha$ being one of the three-components of spin-1 bosons. Here, by mixing the spin-1 antiferromagnetic bosons with spin-0 closed-shell bosons in optical lattices, we observe there are four phases in the low-filling regime, including ferromagnetic phase (FM), unorder insulator (UI), two kinds of superfluid phases (SF), as shown in Fig.~\ref{Na_phase}. Similar to the ferromagnetic case, such as $^{87}{\rm Rb}$, the interspecies interactions between Na and Sr do not influence spin long-range order of spin-1 bosons in the parameters studied here, {\it i.e.} the spin-1 bosons favor spin nematic order both for filling $n=1$ ($n_{\rm Na}=0.5$) and for $n=2$ ($n_{\rm Na}=1$). The physical reason is that the spin-1 bosons favor nematic spin order for filling $n_{\rm Na}$ being odd~\cite{RN44,RN45}. We also examine the local total magnetism of the whole system mixed by $^{23} {\rm Na}$ and $^{84} {\rm Sr}$ in optical lattices, and find that the system favors nonzero magnetism for filling $n=1$ (FM) and zero magnetism for $n=2$ (UI) in the Mott-insulating regime.

\begin{figure}
	\centering
	\includegraphics[width=1\linewidth]{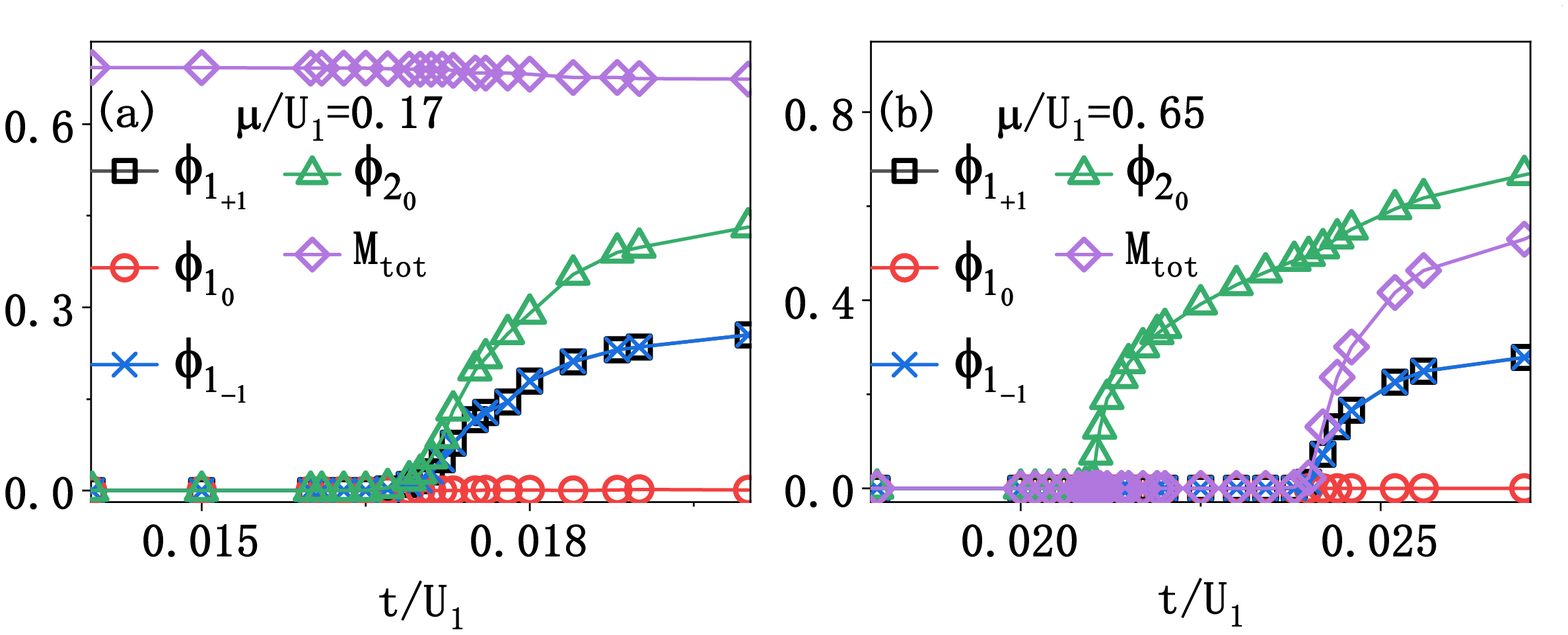}
	\caption{\label{Na} Phase transitions of heteronuclear mixtures of ultracold spin-1 $^{23}$Na and spin-0 $^{84}$Sr bosons in a 3D cubic lattice. We observe a second-order Mott-insulating-superfluid transition for chemical potential $\mu/U_1=0.17$ ($n=1$ in the Mott regime), and a two-step phase transition for  $\mu/U_1=0.65$ ($n=2$ in the Mott regime), {\it i.e.} the Sr atoms delocalized first and then the Na atoms. Here, we choose interactions $U_{12}/U_1=0.4$, $U_{2}/U_1=0.66$, and hopping amplitudes $t \equiv t_{1_\sigma}=t_{2_0}$.}
\end{figure}
With increasing hopping amplitudes, density fluctuations dominate and the system demonstrates a phase transition from the Mott insulator to the superfluid phase, as shown in Fig.~\ref{Na}. As expected, we observe a second-order Mott-insulating-superfluid transition for filling $n=1$. Similar to the mixtures of Rb and Sr, we also observe a two-step Mott-insulating-superfluid phase transition at higher filling, {\it i.e.} the Sr atoms delocalize first, and then the Na atoms, which is a result of the mass imbalance between the Na and Sr atoms. Note here that the spin-1 $^{23}$Na atoms support a nematic-insulating-polar-superfluid phase transition both for local total filling $n=1$ and 2, which is different from the situation with only the spin-1 bosons loaded into optical lattices for even fillings.

\section{conclusions}
In conclusion, we have investigated quantum phases of binary mixtures of spin-1 alkali and spin-0 alkaline-earth bosons loaded into a cubic optical lattice, based on newly developed four-component bosonic dynamical mean-field theory. Complete phase diagrams both with ferromagnetic and antiferromagnetic interactions are obtained. Interestingly, we find that the system demonstrates nonzero magnetic long-range order and a second-order Mott-insulating-superfluid phase transition for filling $n=1$, and a first-order phase transition for $n=2$. In addition, we observe a two-step Mott-insulating-superfluid phase transition, as a result of mass imbalance between alkali and alkaline-earth atoms. We expect the spontaneous spin long-range order of heteronuclear mixtures in optical lattices can be realized and detected using current experimental techniques~\cite{MazurenkoA}.

\begin{acknowledgments}
We acknowledge useful discussions with Liang He, Yu Chen and Hui Zhai. This work was supported by National Natural Science Foundation of China under Grants No. 11304386, and No. 11774428. The work was carried out at National Supercomputer Center
in Tianjin, and the calculations were performed on TianHe-1A.

\end{acknowledgments}

\bibliography{spin1-spin0}

\end{document}